# A compact scalable phase modulator with zero static power consumption for visible integrated photonics

Neil MacFarlane and Firooz Aflatouni

Department of Electrical and Systems Engineering, University of Pennsylvania, Philadelphia, PA 19104

*firooz@seas.upenn.edu

**Optical modulators in the visible regime have far-reaching applications from biophotonics[1,2] to quantum science[3–6]. Implementations of such optical phase modulators on a complementary metal–oxide–semiconductor (CMOS) compatible platform have been mainly limited to utilization of the thermo-optic effect[7,8], liquid crystal technology[9–11], as well as piezo-optomechanical effects[12]. Despite excellent performance, the demonstrations using the thermo-optic effect and liquid crystal technology both suffer from limited modulation speed. Moreover, the demonstrations utilizing piezo-optomechanical effects, require very large footprints due to a weak modulation efficiency. Here, we report the demonstration of the first highly scalable compact CMOS-compatible phase modulator in the visible regime based on altering the refractive index of an indium-tin oxide capacitive stack over a $Si_3N_4$ waveguide through the charge accumulation effect. The implemented modulator achieves a two orders-of-magnitude larger bandwidth compared to thermo-optic and liquid crystal based counterparts and close to 3 orders-of-magnitude higher modulation efficiency with about two orders-of-magnitude smaller footprint compared to piezo-optomechanical modulators. The 50 µm long phase modulator achieves a modulation efficiency, $V_\pi L$, of 0.06 V.cm at a zero static power consumption and a 31 MHz bandwidth at 637.9 nm.**

Integrated photonic devices operating at visible wavelengths have gained significant interest due to the growing applications in biophotonics[1,2] and quantum science[3–6]. Low loss propagation of wavelengths in the visible regime has been demonstrated on numerous material platforms including silicon nitride[4], tantalum pentoxide[13], lithium niobate[14] and titanium dioxide[15]. Silicon nitride, notably, in addition to propagating visible wavelengths has the benefit of being CMOS compatible. Silicon nitride based visible wavelength integrated photonics therefore have the significant advantage of being monolithically integrated with silicon-based electronics as well as silicon based electro-optic components in the visible regime such as photodiodes. Although silicon photonic components operating at infrared wavelengths are both mature and numerous, even the most fundamental silicon nitride components operating at visible wavelengths are less explored. Phase modulation is used ubiquitously in integrated photonic systems for numerous applications such as routing and filtering signals[16], encoding electrical data on optical carriers[17,18], steering emitted beams from optical phased arrays (OPA)[19], and laser phase noise reduction[20]. The mechanisms often used for inducing an optical phase shift include thermo-optic effect[21], the Pockels electro-optic effect[17,18] and the plasma dispersion effect[22].

With the increasing sophistication and complexity of integrated photonic systems and therefore the potential for integrating many phase modulators on a single chip, the power consumption of the chosen phase modulator is an essential consideration. The ideal device would induce a large phase shift with minimal power consumption. The Pockels electro-optic effect, a second-order nonlinear effect that is present in materials that do not possess inversion symmetry of their crystal structure[12], can be used for efficient phase modulation in material platforms such as lithium niobate[17,18], aluminum nitride[23,24] and silicon carbide[25]. However, CMOS compatible materials such crystalline silicon and silicon nitride, do not possess this characteristic intrinsically and therefore produce weak or no phase shift via the Pockels electro-optic effect.

The thermo-optic effect can be utilized to introduce optical phase shifts in most material platforms but at different levels of efficiency. Crystalline silicon for example, has a thermo-optic coefficient of about $1.8 \times 10^{-4}$ 1/K[26], while silicon nitride's thermo-optic phase shifting efficiency is about an order-of-magnitude smaller with a thermo-optic coefficient of $2.45 \times 10^{-5}$ 1/K[27]. Although thermal phase shifting is typically a low-loss technique, it often suffers from a rather high energy consumption, slow time constant (resulting in a low operation speed) and can result in complications due to thermal crosstalk. Furthermore, the generated heat could pose issues for electronic devices co-integrated with the thermo-optic phase modulator.

The plasma dispersion effect can also be used to introduce optical phase shifts. This effect, which has been studied and utilized extensively in crystalline silicon, is based on the Drude-Lorentz model of permittivity[28], where a change in the concentration of holes and electrons causes a change in refractive index and thus a phase shift. By doping the silicon and creating p-n junctions, an applied voltage can modulate the waveguide carrier concentration and induce a phase shift[29]. An alternative architecture is to have an insulating layer, such as silicon dioxide, between two conductive layers (e.g. the oxide layer in a metal-oxide-semiconductor stack), creating a capacitor[30,31]. Applying a voltage across the capacitor creates an accumulation layer of carriers and a change in refractive index. Such a capacitive phase shifting architecture has zero static power consumption and its operation bandwidth is, to the first order, limited only by the resistance and capacitance of the device. This capacitive phase shifting technique, for example, was demonstrated in crystalline silicon for zero static power wavelength tuning of a micro-ring resonator used for PAM-4 modulation[31]. Unlike crystalline silicon, silicon nitride is an insulating material and creating a similar architecture in silicon nitride has not been shown. In general, phase modulation of silicon nitride waveguides is achieved via the thermo-optic effect, however there are demonstrations that achieve modulation using heterogeneous integration of silicon nitride with graphene[32], piezoelectric lead zirconate titanate (PZT)[33], lithium niobate[34], and zinc oxide[35] at infrared wavelengths.

Excellent high-speed phase modulation has been demonstrated using thin film lithium niobate waveguides at visible wavelengths[36], however, such devices are not CMOS compatible and suffer from the low-speed modulation instabilities common in thin film lithium niobate modulators[37]. Silicon nitride as a CMOS compatible material system has been used to implement phase modulators at visible wavelengths using the thermo-optic effect[7,8], integrated liquid crystal technology[9–11], and piezo-optomechanical effects[12]. The efficiency of the thermo-optic approaches is limited due to the low thermo-optic coefficient of silicon nitride. Modulation speed is also limited to the kHz range with demonstrations using a Mach-Zehnder architecture[7] having modulation speeds less than 1 kHz, while a ring resonator architecture[8] showed modulation speeds of 190 kHz. The liquid crystal approach demonstrates strong and low-power phase modulation, but requires complex post-fabrication processes and has modulation speeds that are currently limited to less than 1 kHz[38,39]. Furthermore, the liquid crystal devices typically cannot be DC biased due to potential damage from ionic charge buildup[9]. Using strain induced phase modulation via piezo-optomechanical index tuning of $Si_3N_4$ waveguides, a -3 dB modulation bandwidth of 120 MHz was achieved at 700 nm wavelength. However, despite excellent demonstration, due to the relatively weak index strain-optic effect in $Si_3N_4$, a rather large device with a footprint of $10^5$ μm$^2$ was used and a limited modulation efficiency of $V_\pi L$ ~ 50 V.cm was achieved[12].

Here, we report the demonstration of the first CMOS-compatible $Si_3N_4$ phase modulator in the visible regime with a zero static power consumption. Using an indium-tin oxide (ITO) based capacitive stack on a silicon nitride waveguide, the effective index of the guided mode is changed through the charge accumulation effect modulating the phase of the optical wave. The implemented ITO capacitive modulator achieves a two orders-of-magnitude larger bandwidth compared to thermo-optic and liquid crystal based designs at a zero static power consumption and, compared to piezo-optomechanical modulators, achieves an almost 3 orders-of-magnitude smaller $V_\pi L$ and occupies over two orders-of-magnitude smaller footprint at a competitive modulation bandwidth. A Mach Zehnder modulator (MZM) was formed using the ITO capacitive phase modulators. The 50 µm long phase modulator (fabricated within a footprint of 750 µm$^2$) achieves a modulation efficiency, $V_\pi L$, of 0.06 V.cm and a 31 MHz bandwidth at 637.9 nm. The bandwidth of the implemented devices is mainly limited only by the RC time constraints and could be optimized for a target application given the tradeoff between bandwidth, modulation efficiency, and optical insertion loss. This design represents a highly efficient phase shifting approach with a zero static power consumption for visible wavelength integrated photonics.

**Results**

**ITO capacitive modulator in visible regime**

Transparent conductive oxides (TCO) are a class of materials that have the unique property of being both optically transparent and electrically conductive. This is a key feature in implementation of the proposed capacitive modulator. Indium tin oxide is a particularly common TCO that is used extensively in the solar cell industry and is compatible with CMOS technology. The structure and cross section of the implemented ITO capacitive phase modulator are shown in Fig. 1a and 1b respectively. The devices are fabricated on a 4-inch silicon wafer with 3 µm of thermal silicon dioxide on the surface to act as a lower cladding material. Silicon nitride films are deposited using low-pressure chemical vapor deposition (LPCVD)[40,41]. Next, the ITO capacitive stack is formed by first depositing a 30 nm thick ITO film (serving as the capacitor bottom plate) on a 200 nm thick $Si_3N_4$ waveguide using a DC sputtering process[42]. A 20 nm thick $HfO_2$ dielectric layer is deposited on the ITO layer using atomic layer deposition. Another ITO layer, serving as the capacitor top plate, is deposited on the dielectric layer, forming the ITO-$HfO_2$-ITO parallel plate capacitor over the $Si_3N_4$ waveguide. Lastly gold contacts are deposited using electron-beam evaporation. Gold is the chosen metal because it has a similar work function to ITO, which results in ohmic contact to ensure linear performance of the ITO modulator. More details of the fabrication process can be found in the Methods section. Note that the chosen film thickness was 200 nm, which enables overlap of the optical mode with the ITO capacitor stack, strong effective index modulation, and therefore efficient phase shifting. Figure 1b also shows the optical mode within the phase modulator, where the wave is confined within $Si_3N_4$ and ITO-$HfO_2$-ITO layers.

When a voltage, $V_M$, is applied across the terminals of the capacitive phase modulator, charge accumulation within the ITO layers of the ITO-$HfO_2$-ITO capacitor results in a change in the carrier concentration and profile. The overlap between the optical mode within the $Si_3N_4$ waveguide, ITO films, and the insulator region with the charge distribution profile can be used to estimate the change in the effective refractive index that results in modulation of the optical wave.

The carrier density of the accumulation layer is a function of the capacitance and the applied voltage. Therefore, starting with a low carrier concentration allows for a large change in carrier concentration for a given applied voltage. However, low carrier concentration of the ITO films can also result in higher resistance (limiting the operation bandwidth). We chose to deposit the ITO films with a low carrier concentration and increased the carrier concentration through controlled annealing and $O_2$ plasma treatment[42], which also affect the bandwidth and modulation efficiency of the devices.

The first step in device characterization is to find the propagation loss of the $Si_3N_4$ waveguide as well as the intrinsic propagation loss of the ITO capacitive phase modulators. The transmission through $Si_3N_4$ waveguides with varying lengths is measured at 637.9 nm and average propagation losses of 3.4 dB/cm, 1.2 dB/cm and 0.9 dB/cm for waveguide widths of 500 nm, 800 nm and 1500 nm were measured, respectively, which are comparable to those reported from the

AMF foundry silicon nitride films[7]. The cutback measurements for the smallest and largest width waveguides are shown in Figure 1c. The propagation loss of the silicon nitride waveguide with ITO capacitor stack is also measured, where the length of the silicon nitride waveguide is kept constant at 1 mm and the length of the ITO capacitor stack is varied from 50 µm to 900 µm. The measurement results for the propagation losses of the ITO capacitive modulator losses with no annealing, with 200º C annealing and with 300º C annealing are shown in Fig. 1d. The propagation losses with no annealing and with annealing at 200 ºC stays relatively constant at ~20 dB/mm and then increases to ~80 dB/mm after annealing at 300º C. The increase in loss after annealing at 300º C is in agreement with previous reports that annealing ITO films increases the carrier concentration, which would also increase the absorption loss[42]. We anticipate that the high loss is due to a combination of absorption in the ITO films and scattering at the ITO/$Si_3N_4$ waveguide interfaces. With further optimization of the fabrication process, the propagation loss due to scattering should be decreased.

Figure 1e shows the measured refractive index of the deposited ITO films (using ellipsometry) as a function of wavelength, and the fit to the data based on the Drude-Lorentz model of permittivity, which was then used to estimate the carrier concentration. As expected, lower sputter deposition power results in lower as-deposited film carrier concentration[42]. More details on the refractive index measurement are included in the Methods section.

The phase modulation based on altering the partial mode of a $Si_3N_4$ waveguide through charge accumulation is modelled and simulated. The simulated ITO charge accumulation and resulting effective index change are plotted as a function of the voltage across the capacitive modulator, $V_M$, in Figure 1f. The effective index changes by about $5.5 \times 10^{-4}$ refractive index units over the simulated voltage range of -10 V to 10 V, which confirms the viability of the proposed approach for optical phase shifting in the visible regime using silicon nitride waveguides. The details of the simulation can be found in the Methods section.

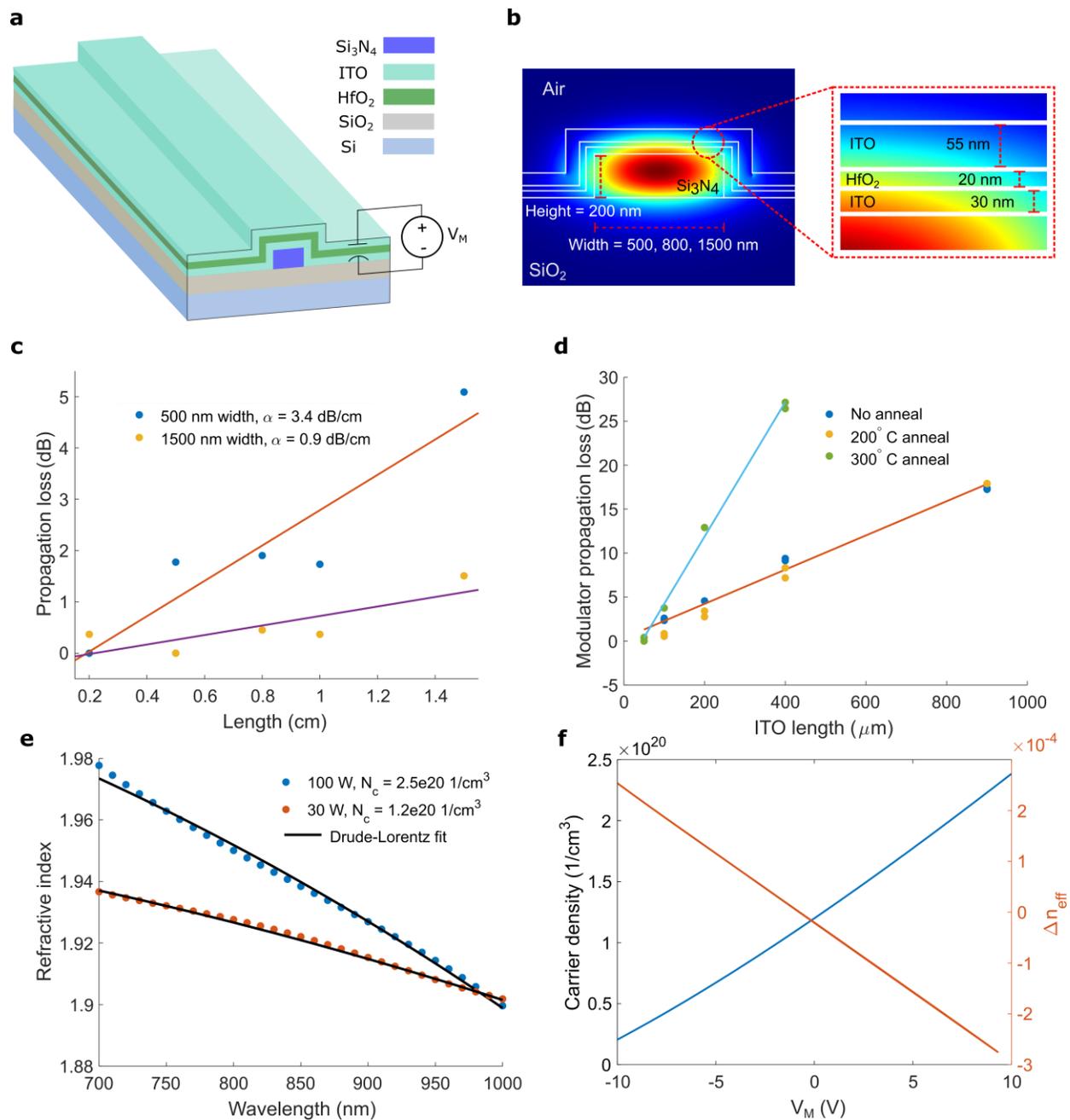

**Fig. 1 | Structure and passive characterization of ITO capacitive phase modulators. a.** ITO capacitive phase modulator device showing the capacitive stack on $Si_3N_4$. **b.** Cross section and simulated mode profile of ITO-$HfO_2$-ITO stack on $Si_3N_4$ waveguide. **c.** Measured propagation loss of silicon nitride waveguides. **d.** Measured propagation loss of 1500 nm wide silicon nitride waveguides with ITO capacitor stack. **e.** Refractive index data of ITO films extracted from ellipsometry measurements. **f.** Simulation of ITO carrier concentration and resulting change in the effective index of the guided mode of the waveguide structure (ITO capacitive stack on $Si_3N_4$).

## Low-frequency electro-optic response of ITO capacitive phase modulators

The ITO phase modulators are characterized by measuring the output intensity of a Mach-Zehnder modulator (MZM) driven in a push-pull configuration. The structure of the fabricated balanced MZM is shown in Fig. 2a. The output of a QPhotonics diode laser emitting up to 40 mW at 637.9 nm is coupled into the chip using the designed lensed grating coupler and is routed to the input 1x2 multimode interference (MMI) coupler (serving as a splitter) using a 500 nm wide $Si_3N_4$ waveguide. An ITO capacitive phase modulator is placed on each arm of the Mach-Zehnder interferometer (MZI) to match the loss of two arms. The same MMI device serving as a 2x1 combiner is used to combine the outputs of the phase modulators and the MZM output is coupled out of the chip using a grating coupler. The scanning electron micrograph of the grating coupler and the optical micrograph of the MMI coupler are shown in Figure 2a(i) and 2a(ii), respectively. More details on the grating coupler and the MMI coupler designs are included in the Methods section. A top-down scanning electron microscope image and the cross-section of the ITO capacitive phase modulators are shown in Figs. 2a(iii) and 2a(iv), respectively. An optical micrograph of the fabricated MZM device is shown in Fig. 2b.

Figure 3a shows the measurement setup used to characterize the low frequency electro-optic response of the modulator, where a 20 $V_{pp}$ linear ramp voltage with a period of 33.3 ms was applied to the MZM and the measured corresponding optical response was measured as shown in Fig. 3b.

To study the effect of each fabrication step on the device performance, the low frequency electro-optic response of the modulator was measured after four different process steps; before annealing, after 200º C annealing, after 300º C annealing, and finally after 300º C annealing with $O_2$ plasma treatment. The output intensity as a function of applied voltage after each of the processing steps is shown in Figs. 3c-f, where the $V_\pi L$ of the ITO modulators reduces from 0.5 V·cm before annealing to 0.06 V·cm after annealing at 300º C and treating with $O_2$ plasma. This is caused by the increase in the carrier concentration of the ITO films, which results in a larger change in the waveguide effective index as a function of the applied voltage, significantly reducing the $V_\pi L$ of the modulator at the cost of increased propagation loss (due to an increase in the carrier absorption). The change in the effective index after each of the four processing steps is estimated from the measurements, which is shown in Fig. 3g. While there is a small difference between the unannealed and 200º C annealed cases, annealing at 300º C increases the effective index change at 10 V from 3e-4 to 0.7e-3, which further increases to 2.3e-3 after $O_2$ plasma treatment. This is almost an order-of-magnitude increase in waveguide effective index change compared to the effective index change measured before annealing. Note that the refractive index change for the unannealed sample is in close agreement with the simulation results in Fig. 1f.

The low-speed electro-optic response measurements shows that the compact ITO capacitive phase modulator features a low $V_\pi L$, while consuming zero static power. More details of the measurement are summarized in the Methods section.

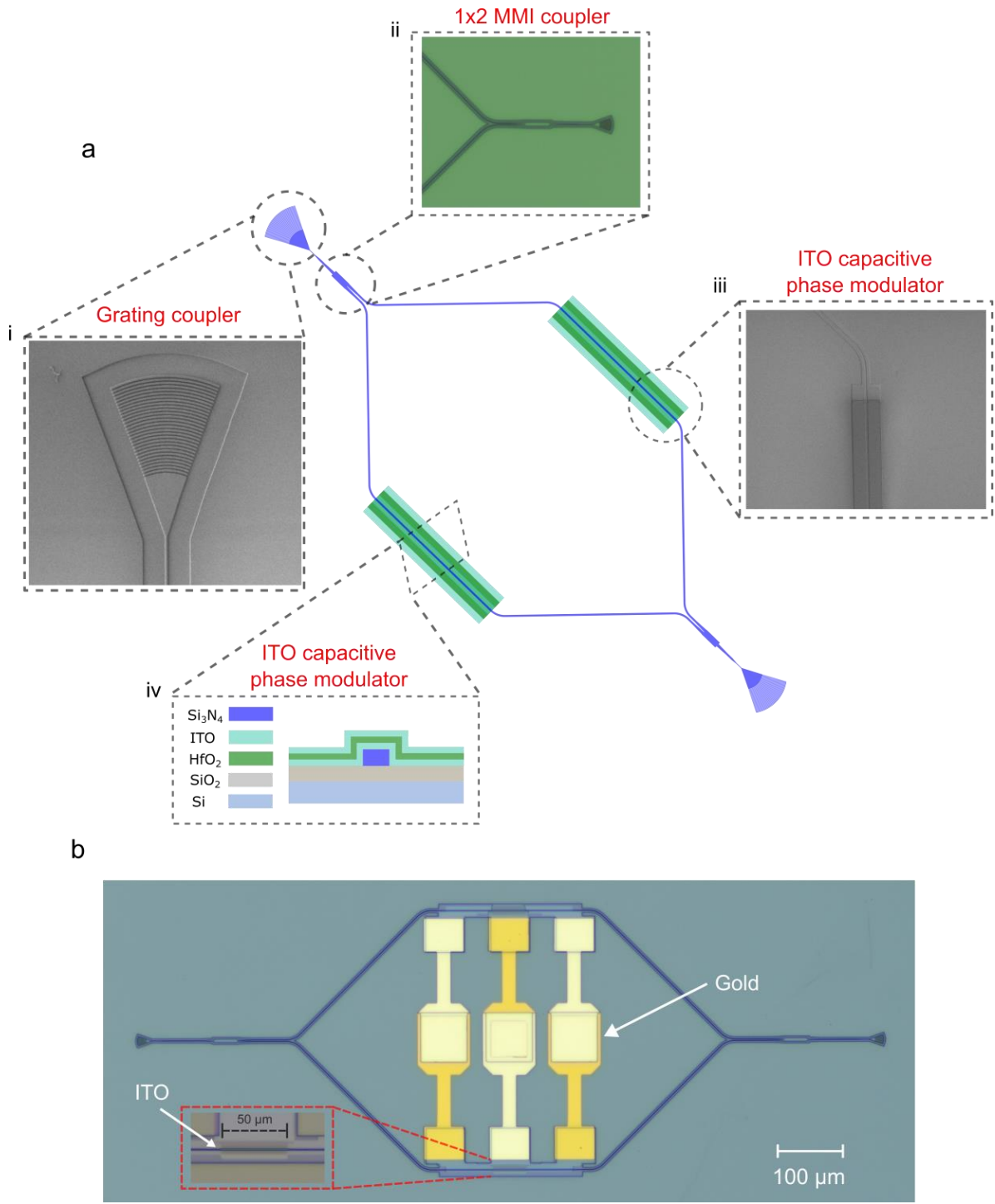

**Fig. 2 | MZM architecture. a.** ITO capacitor Mach-Zehnder Modulator, **i.** Scanning electron micrograph of grating coupler. **ii.** Optical micrograph of MMI coupler. **iii.** Scanning electron micrograph of ITO capacitive phase modulator. **iv.** Cross section of ITO capacitive phase modulator. **b.** Optical micrograph of fabricated ITO capacitor Mach-Zehnder modulator. (**Inset.** 50 µm length ITO capacitive phase modulator).

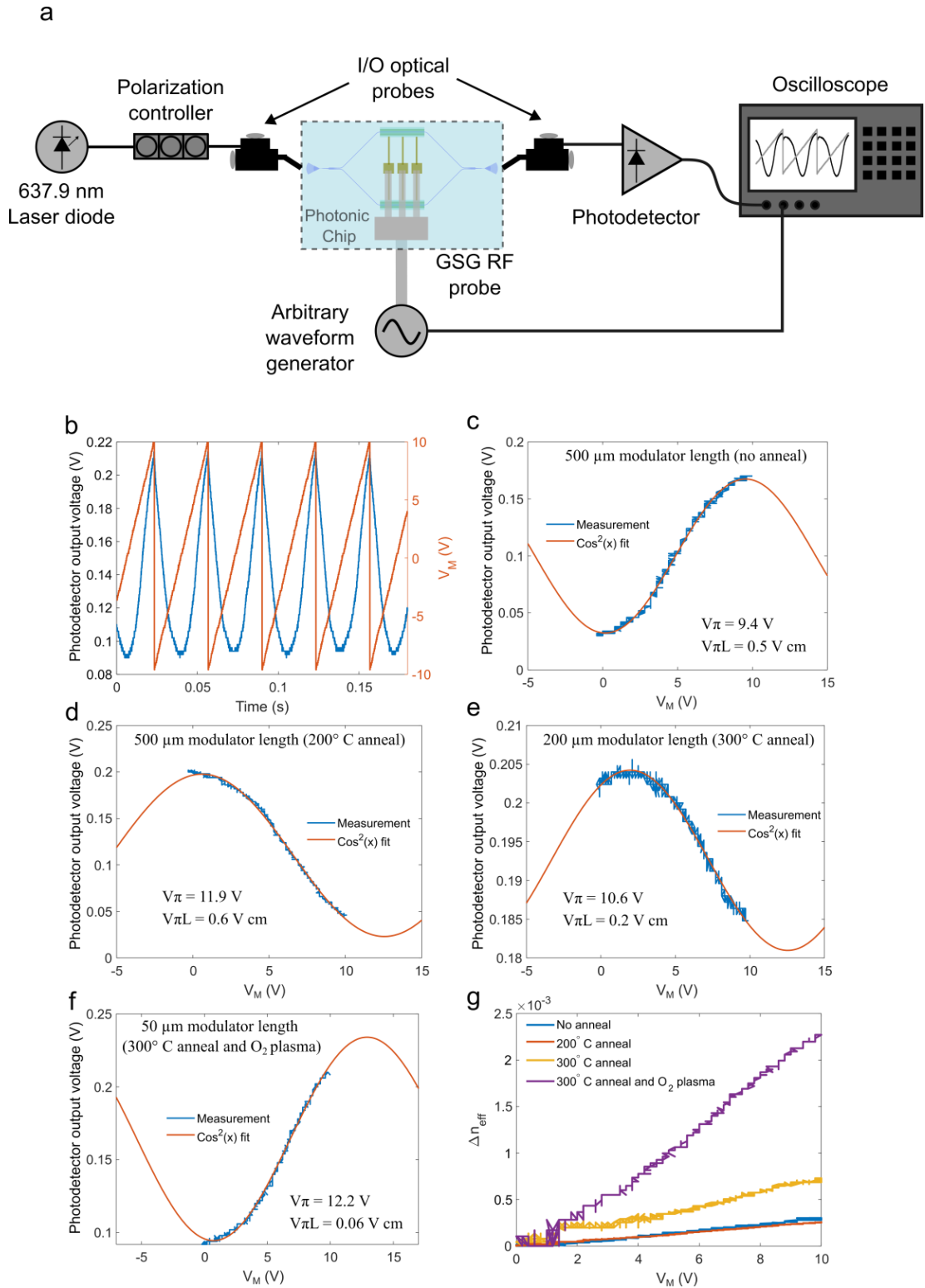

**Fig. 3 | Low-frequency electro-optic response of ITO capacitive modulators. a.** Measurement setup for low-frequency electro-optic response of the ITO capacitive modulator. **b.** Applied voltage ramp and electro-optic response of the ITO capacitive modulator. **c-f.** $Cos(x)^2$ fit of $V_\pi$ measurement for ITO capacitive phase modulator with no annealing, 200 ºC annealing, 300 ºC annealing, and 300 ºC annealing and $O_2$ plasma treatment, respectively. **e.** Change in effective index as a function of applied voltage extracted from measurements in **c-f**.

**High-speed electro-optic response of ITO capacitive phase modulators**

Modulation bandwidth, often limited by the RC time constant, is another important consideration for an electro-optic modulator. The modulation bandwidth of the ITO capacitive phase modulator, mainly limited by the device capacitance and series resistance, was measured for different device lengths ranging from 50 µm to 300 µm. Moreover, the bandwidth dependence on annealing process, which effectively changes the carrier concentrations and as a result the device series resistance, was experimentally studied by measuring the electro-optic response of the modulators before and after annealing at 200º C and 300º C. The measurement setup for high-speed electro-optic characterization of the ITO capacitive modulators is shown in Fig. 4a, which closely resembles the setup used for low frequency ITO capacitive modulator measurements. To find the modulation bandwidth the amplitude of the electro-optic response at the applied RF frequency is monitored.

The measured electro-optic response of a 50 µm long ITO capacitive modulator as a function of modulation frequency for different annealing temperatures is shown in Fig. 4b. As expected, annealing decreases the resistance of the ITO films. As a result, the -3 dB bandwidth of the device increases from less than 50 kHz in the case of no annealing to about 31 MHz after annealing at 300º C. The modulator response for different phase modulator lengths (after annealing at 300 ºC and $O_2$ plasma treatment) is also measured, which is shown in Fig. 4c, where a -3 dB bandwidth of about 3 MHz for the 300 µm long device increases to about 31 MHz for the 50 µm long device. We note that the bandwidth does not increase linearly with the decrease in capacitance from the 300 µm length to 50 µm modulators, which could indicate that in addition to the decrease in capacitance, the shorter modulator also results in a decrease in resistance. We confirmed this by distributed modelling the equivalent electrical circuit of the 300 µm as smaller RC sections. Figure 4d shows the curve fitting used to estimate the bandwidth of the 50 µm long device from the measured electro-optic response. More details on the measurement process are included in the Methods section.

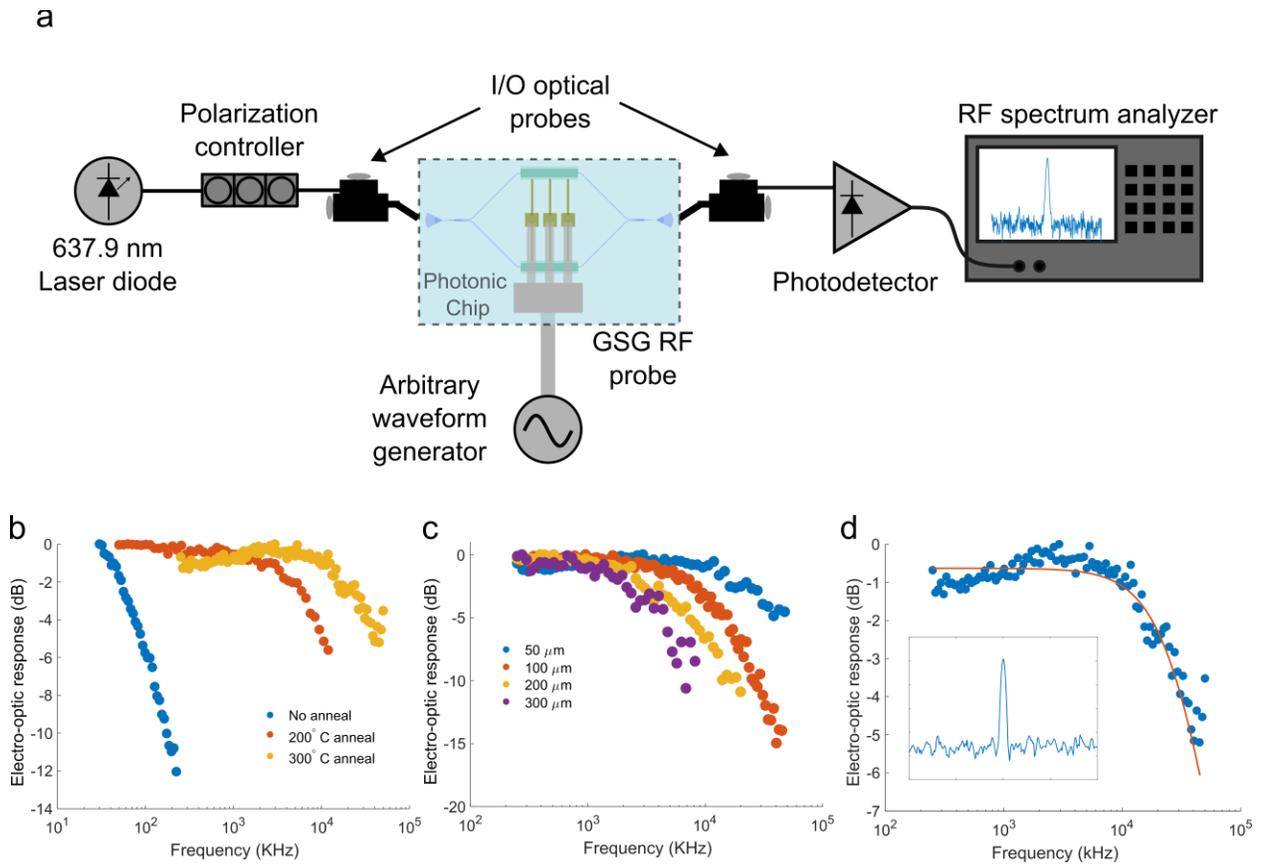

**Fig. 4 | Bandwidth characterization of ITO capacitive phase modulators. a.** Experimental setup for high-speed characterization. **b.** Measured electro-optic response of a 50 µm ITO capacitive modulators before and after annealing. **c.** Measured electro-optic response of 300º C annealed ITO capacitive modulators of various lengths. **d.** Measured electro-optic response of a 50 µm length modulator and fit used to estimate the -3 dB bandwidth. (**Inset:** example generated sideband at 250kHz modulation frequency).

| | Phase modulation Mechanism | Waveguide Material | Structure | λ (nm) | Length (mm) | Area (µm²) | Loss (dB) | $V_\pi L$ (V.cm) | $P_\pi$ (mW) | Bandwidth (MHz) | Drive Voltage |
|---|---|---|---|---|---|---|---|---|---|---|---|
| 7 | Thermal (with trenches) | $Si_3N_4$ | MZM | 561 | 1.5 | 6300 | 4.8 | N/A | 1.22 | 0.0006 | DC/AC |
| 8 | Thermal | $Si_3N_4$ | Ring | 530 | N/A | 315 | 0.87 | N/A | 0.85 | 0.190 | DC/AC |
| 9 | Liquid crystal | $Si_3N_4$ | MZM | 632.8 | 0.5 | 2500 | 0.5 | N/A | N/A | N/A | AC |
| 12 | Piezo-optomechanical strain-tuning | $Si_3N_4$ | MZM | 700 | 10 | $10^5$ | 3.5 | 50 | N/A | 120 | DC/AC |
| This work | Charge accumulation (ITO capacitor) | $Si_3N_4$ | MZM | 637.9 | 0.05 | 750 | 4 | 0.06 | 0 | 31 | DC/AC |

**Table 1 | Comparison of our work with other demonstrations of CMOS compatible visible wavelength phase modulators.**

**Discussion**

A novel method for achieving efficient phase modulation at visible wavelengths has been demonstrated. Using a repeatable and high yield fabrication process, an ITO based capacitive phase modulator on silicon nitride waveguides was demonstrated with modulation bandwidths up to 31 MHz, and $V_\pi L$ values as low as 0.06 V.cm at a zero static power consumption. Measurements show that a combination of annealing and $O_2$ plasma treatment increases the modulation bandwidths and reduces the $V_\pi L$. The performance of the devices implemented in this work is compared with that of other CMOS compatible phase modulators in Table 1.

We have demonstrated a capacitive modulator that compared to thermo-optic and liquid crystal based modulators in the visible regime, achieves a two orders-of-magnitude larger bandwidth while consumption a zero static power. Furthermore, while the bandwidth of the implemented modulator is competitive with the piezo-optomechanical design, it achieves an almost 3 orders-of-magnitude smaller $V_\pi L$ and has an over two orders-of-magnitude smaller footprint.

Not that, there are a number of ways to further increase the bandwidth of the implemented phase modulator. One way, to the first order, is to decrease the capacitance by decreasing the width of the ITO layer, which is currently 15 μm and could be reduced to close to the $Si_3N_4$ waveguide width using electron-beam lithography to pattern the ITO layer resulting in a smaller device area and capacitance. Other ways to enhance the device bandwidth include increasing the thickness of the dielectric layer or using a dielectric material with a lower dielectric constant.

The overall resistance of the device could be reduced by decreasing the contact resistance between the ITO and gold contacts. Potential optimizations include selectively exposing the ITO to $O_2$ plasma treatment in the contact regions, which effectively dopes the ITO films and increases conductivity[42], or using ITO films with higher as deposited carrier concentration in the contact area.

The 4 dB insertion loss of the implemented phase modulators is due to a combination of scattering loss and absorption. While the absorption loss is somewhat unavoidable due to the free carriers in ITO, the scattering loss may be reduced through fabrication optimization. Scattering losses could be reduced by depositing a silicon dioxide cladding on the $Si_3N_4$ waveguides followed by chemical mechanical polishing down to the top of the $Si_3N_4$ waveguides before fabricating the ITO-$HfO_2$-ITO capacitors. In this way, there would still be strong interaction of the optical mode with the ITO at the surface of the waveguide, but the cladding layer would increase the confinement of the optical mode near the sidewalls and therefore reduce the scattering losses. Additionally, the scattering loss from sidewall roughness, induced during the etch process in $Si_3N_4$ waveguides, could be reduced with fabrication optimization. For example, one could explore cold development of ZEP520A electron-beam lithography resist, which has been shown to be an effective method of reducing line edge roughness of the developed patterns[43].

Note that, unlike the liquid crystal phase modulators that cannot be DC biased (to avoid damage from ionic charge buildup), our approach can be used with DC or AC drive voltages. Lastly, besides the zero static power consumption of our devices compared to the demonstrated thermal modulators with reported $P_\pi$ values of 1.22 mW at 561 nm wavelength[7] and 0.85 mW at 530 nm wavelength[8], an integrated photonic system utilizing multiple ITO capacitive phase modulators (instead of thermal modulators) does not suffer from thermal crosstalk issues.

Although silicon nitride was the material of choice, our design allows for versatility in terms of the material platform, with the only constraint being the ability to propagate visible wavelengths without significant absorption. Therefore, the implemented device could be realized using other insulator materials, particularly, low temperature deposition materials such as $Ta_2O_5$ and $NbTaOx$, which are two metal oxide materials shown to be viable alternatives to $Si_3N_4$ at infrared[44,45] and visible wavelengths[13]. Given the low temperature processing of the ITO phase modulators, using a waveguide material platform allowing for a low temperature deposition process enables seamless back end of the line integration with CMOS electronics and foundry-based silicon photonic chips.

One potential application for this technology is in optical phased arrays in visible regime, where using our ITO based capacitive phase shifting approach for beam steering, improvements on power efficiency and steering speed can be achieved compared to previous demonstrations of visible OPAs[11,46].

**References**

1. Sacher, W. D. *et al.* Visible-light silicon nitride waveguide devices and implantable neurophotonic probes on thinned 200 mm silicon wafers. *Opt Express* **27**, 37400–37418 (2019).
2. Kohler, D. *et al.* Biophotonic sensors with integrated Si3N4-organic hybrid (SiNOH) lasers for point-of-care diagnostics. *Light Sci. Appl.* **10**, 64 (2021).
3. Spektor, G. *et al.* Universal visible emitters in nanoscale integrated photonics. *Optica* **10**, 871–879 (2023).
4. Chauhan, N. *et al.* Ultra-low loss visible light waveguides for integrated atomic, molecular, and quantum photonics. *Opt Express* **30**, 6960–6969 (2022).
5. Niffenegger, R. J. *et al.* Integrated multi-wavelength control of an ion qubit. *Nature* **586**, 538–542 (2020).
6. Saha, U. *et al.* Routing Single Photons from a Trapped Ion Using a Photonic Integrated Circuit. *Phys Rev Appl* **19**, 034001 (2023).
7. Yong, Z. *et al.* Power-efficient silicon nitride thermo-optic phase shifters for visible light. *Opt Express* **30**, 7225–7237 (2022).
8. Liang, G. *et al.* Robust, efficient, micrometre-scale phase modulators at visible wavelengths. *Nat. Photonics* **15**, 908–913 (2021).
9. Notaros, M. *et al.* Integrated visible-light liquid-crystal-based phase modulators. *Opt Express* **30**, 13790–13801 (2022).
10. Notaros, M., Coleto, A. G., Raval, M. & Notaros, J. Integrated liquid-crystal-based variable-tap devices for visible-light amplitude modulation. *Opt Lett* **49**, 1041–1044 (2024).
11. Notaros, M., DeSantis, D. M., Raval, M. & Notaros, J. Liquid-crystal-based visible-light integrated optical phased arrays and application to underwater communications. *Opt Lett* **48**, 5269–5272 (2023).
12. Dong, M. *et al.* High-speed programmable photonic circuits in a cryogenically compatible, visible–near-infrared 200 mm CMOS architecture. *Nat. Photonics* **16**, 59–65 (2022).
13. Irvine, D. A. *et al.* Propagation Losses of Sputtered Oxide Waveguides in the Visible Range. in *CLEO 2023* SM2H.5 (Optica Publishing Group, 2023). doi:10.1364/CLEO_SI.2023.SM2H.5.
14. Desiatov, B., Shams-Ansari, A., Zhang, M., Wang, C. & Lončar, M. Ultra-low-loss integrated visible photonics using thin-film lithium niobate. *Optica* **6**, 380–384 (2019).
15. Choy, J. T. *et al.* Integrated TiO2 resonators for visible photonics. *Opt Lett* **37**, 539–541 (2012).
16. Bogaerts, W. *et al.* Programmable photonic circuits. *Nature* **586**, 207–216 (2020).
17. Wang, C. *et al.* Integrated lithium niobate electro-optic modulators operating at CMOS-compatible voltages. *Nature* **562**, 101–104 (2018).
18. Mercante, A. J. *et al.* Thin film lithium niobate electro-optic modulator with terahertz operating bandwidth. *Opt Express* **26**, 14810–14816 (2018).
19. Aflatouni, F., Abiri, B., Rekhi, A. & Hajimiri, A. Nanophotonic projection system. *Opt Express* **23**, 21012–21022 (2015).
20. Idjadi, M. H. & Aflatouni, F. Nanophotonic phase noise filter in silicon. *Nat. Photonics* **14**, 234–239 (2020).

21. Harris, N. C. *et al.* Efficient, compact and low loss thermo-optic phase shifter in silicon. *Opt Express* **22**, 10487–10493 (2014).
22. Reed, G. T., Mashanovich, G., Gardes, F. Y. & Thomson, D. J. Silicon optical modulators. *Nat. Photonics* **4**, 518–526 (2010).
23. Liu, S., Xu, K., Song, Q., Cheng, Z. & Tsang, H. K. Design of Mid-Infrared Electro-Optic Modulators Based on Aluminum Nitride Waveguides. *J. Light. Technol.* **34**, 3837–3842 (2016).
24. Zhu, S. & Lo, G.-Q. Aluminum nitride electro-optic phase shifter for backend integration on silicon. *Opt Express* **24**, 12501–12506 (2016).
25. Powell, K. *et al.* Integrated silicon carbide electro-optic modulator. *Nat. Commun.* **13**, 1851 (2022).
26. Komma, J., Schwarz, C., Hofmann, G., Heinert, D. & Nawrodt, R. Thermo-optic coefficient of silicon at 1550 nm and cryogenic temperatures. *Appl. Phys. Lett.* **101**, 041905 (2012).
27. Arbabi, A. & Goddard, L. L. Measurements of the refractive indices and thermo-optic coefficients of Si3N4 and SiOx using microring resonances. *Opt Lett* **38**, 3878–3881 (2013).
28. Soref, R. & Bennett, B. Electrooptical effects in silicon. *IEEE J. Quantum Electron.* **23**, 123–129 (1987).
29. Gardes, F. Y., Reed, G. T., Emerson, N. G. & Png, C. E. A sub-micron depletion-type photonic modulator in Silicon On Insulator. *Opt Express* **13**, 8845–8854 (2005).
30. Liao, L. *et al.* High speed silicon Mach-Zehnder modulator. *Opt Express* **13**, 3129–3135 (2005).
31. Omirzakhov, K., Pirmoradi, A., Hao, H. & Aflatouni, F. Monolithic optical PAM-4 transmitter with autonomous carrier tracking. *Opt Express* **32**, 2894–2905 (2024).
32. Phare, C. T., Daniel Lee, Y.-H., Cardenas, J. & Lipson, M. Graphene electro-optic modulator with 30 GHz bandwidth. *Nat. Photonics* **9**, 511–514 (2015).
33. Alexander, K. *et al.* Nanophotonic Pockels modulators on a silicon nitride platform. *Nat. Commun.* **9**, 3444 (2018).
34. Nelan, S. *et al.* Compact thin film lithium niobate folded intensity modulator using a waveguide crossing. *Opt Express* **30**, 9193–9207 (2022).
35. Hermans, A. *et al.* Integrated silicon nitride electro-optic modulators with atomic layer deposited overlays. *Opt Lett* **44**, 1112–1115 (2019).
36. Renaud, D. *et al.* Sub-1 Volt and high-bandwidth visible to near-infrared electro-optic modulators. *Nat. Commun.* **14**, 1496 (2023).
37. Zhang, M., Wang, C., Kharel, P., Zhu, D. & Lončar, M. Integrated lithium niobate electro-optic modulators: when performance meets scalability. *Optica* **8**, 652–667 (2021).
38. Xing, Y. *et al.* Digitally Controlled Phase Shifter Using an SOI Slot Waveguide With Liquid Crystal Infiltration. *IEEE Photonics Technol. Lett.* **27**, 1269–1272 (2015).
39. Atsumi, Y., Watabe, K., Uda, N., Miura, N. & Sakakibara, Y. Initial alignment control technique using on-chip groove arrays for liquid crystal hybrid silicon optical phase shifters. *Opt Express* **27**, 8756–8767 (2019).
40. Ji, X. *et al.* Ultra-low-loss on-chip resonators with sub-milliwatt parametric oscillation threshold. *Optica* **4**, 619–624 (2017).
41. Ji, X., Roberts, S., Corato-Zanarella, M. & Lipson, M. Methods to achieve ultra-high quality factor silicon nitride resonators. *APL Photonics* **6**, 071101 (2021).
42. Ma, Z., Li, Z., Liu, K., Ye, C. & Sorger, V. J. Indium-Tin-Oxide for High-performance Electro-optic Modulation. *Nanophotonics* **4**, 198–213 (2015).
43. Ocola, L. E. & Stein, A. Effect of cold development on improvement in electron-beam nanopatterning resolution and line roughness. *J. Vac. Sci. Technol. B Microelectron. Nanometer Struct. Process. Meas. Phenom.* **24**, 3061–3065 (2006).

44. MacFarlane, N., Schreyer-Miller, A., Foster, M. A., Houck, W. D. & Foster, A. C. Niobium-tantalum oxide as a material platform for linear and nonlinear integrated photonics. *Opt Express* **30**, 42155–42167 (2022).
45. Jung, H. *et al.* Tantala Kerr nonlinear integrated photonics. *Optica* **8**, 811–817 (2021).
46. Shin, M. C. *et al.* Chip-scale blue light phased array. *Opt Lett* **45**, 1934–1937 (2020).
47. Soldano, L. B. & Pennings, E. C. M. Optical multi-mode interference devices based on self-imaging: principles and applications. *J. Light. Technol.* **13**, 615–627 (1995).
48. Hong, J., Spring, A. M., Qiu, F. & Yokoyama, S. A high efficiency silicon nitride waveguide grating coupler with a multilayer bottom reflector. *Sci. Rep.* **9**, 12988 (2019).
49. Almeida, V. R., Panepucci, R. R. & Lipson, M. Nanotaper for compact mode conversion. *Opt Lett* **28**, 1302–1304 (2003).

## Methods

### Device Fabrication

The fabrication process begins with a 4-inch silicon wafer with 3 μm of thermal silicon dioxide on the surface to act as a lower cladding material. The silicon nitride films are deposited using low-pressure chemical vapor deposition (LPCVD). We deposit our films using a Sandvik Furnace at 850° C with 50 sccm of ammonia ($NH_3$) and 150 sccm of dichlorosilane (DCS) gas flows. Following deposition, the films are annealed at 1050° C to further reduce absorption losses from hydrogen bonding[41]. This recipe has a deposition rate of 4.2 nm/min. The chosen film thickness was 200 nm. The silicon nitride integrated photonic structures were patterned using the Raith EBPG5200+ electron-beam lithography tool and high-resolution ZEP 520A positive electron-beam resist. The patterns were transferred from the resist to the silicon nitride films using an Oxford 80 reactive ion etcher with $CF_4$ based etching chemistry. The remaining resist is stripped using oxygen plasma etching. Photolithography was performed using the Heidelberg DWL 66+ Laser Writer to open windows for the ITO and metal layers to be deposited and lifted off. The deposited ITO layers in the stack had thicknesses of 30 nm and 55 nm and were deposited using a Lesker PVD75 DC/RF Sputterer with a sputtering power of 30 W and pressure of 7 mTorr. A 20 nm thick $HfO_2$ layer was deposited using the Cambridge Nanotech S200 atomic layer deposition tool. In order to probe the capacitor, 200 nm thick gold metal pads in a ground-signal-ground (GSG) configuration and routing lines are deposited using the Lesker PVD75 E-Beam Evaporator.

**Multimode interference coupler**

Multimode interference (MMI) couplers were used as 50/50 splitters in the MZM devices. MMI splitters consist of a multimode waveguide slab that is probed with one input and two output single mode waveguides. The design of the MMIs began with using the analytical approach described by Soldano and Pennings[47]. First, the length $L_\pi$ (i.e., the propagation length at which the fundamental and first-order modes of the MMI slab are out of phase by $\pi$) was analytically calculated using the equation $L_\pi = \frac{\pi}{\beta_0 - \beta_1}$, where values for the longitudinal wave vectors, $\beta_0$ and $\beta_1$, were found through a Lumerical simulation of the MMI slab waveguide modes. Using restricted interference[47], the single input waveguide is placed at the center of the multimode waveguide and on the opposite side the two output waveguides are placed symmetrically about the center of the multimode waveguide. With this architecture, the analytical length of the MMI for 50/50 splitting is equal to $\frac{3L_\pi}{8}$. Using Lumerical's Eigenmode Expansion Solver with the analytical length as a starting value, the optimized length of the 5 µm wide MMI was set to 35.5 µm for low insertion loss and a 50/50 splitting ratio.

**Visible wavelength grating coupler**

Coupling to the photonic chip is realized using grating couplers designed for the target wavelength of 637.9 nm. The grating couplers are designed and simulated using Lumerical's Finite Difference Time Domain (FDTD) solver. The Bragg condition for the grating coupler was calculated[48], and Lumerical's mode solver was used to find the effective index and grating period of the grating coupler. To reduce the fabrication complexity, only one etch step was performed, that is, the grating coupler tooth has the same height as the waveguides (200 nm) and the trench is etched down 200 nm to the top of the $SiO_2$ layer. The duty cycle and the emission angle were optimized for maximum emitted power using 2D FDTD simulations, resulting in an optimum duty cycle and emission angle of 0.65 and 13º, respectively.

The 2D simulated design was converted to a lensed grating coupler by combining the grating structure and the waveguide taper, resulting in a smaller device footprint[48]. The final grating period and duty cycle of the implemented lensed grating coupler were 464 nm and 0.65, respectively. The total footprint of the final design was 326 µm$^2$. The measured coupling losses of the fabricated lensed grating couplers used in the modulator devices were 10 dB at 637.9 nm. The coupling efficiency of the grating coupler could be further improved by utilizing bottom reflector stacks[48] or exploring other input coupling mechanisms such as edge-coupling using lensed fibers and inverse taper waveguide couplers[49].

**Phase modulation**

To model the ITO capacitive phase modulator, first, the charge accumulation as a function of applied voltage was simulated using the Lumerical CHARGE solver. In this simulation, ITO is modelled as an n-type doped semiconductor. Hafnium dioxide ($HfO_2$) is the dielectric layer for the ITO capacitor, which was chosen for its high dielectric constant and large breakdown voltage compared to silicon dioxide. The charge accumulation results, which are shown in Figure 1f, are then used to simulate the change in the waveguide effective index as a function of applied voltage using the Lumerical MODE solver. The silicon nitride waveguide in this simulation has dimensions of 500 x 200 nm. The effective index of refraction of the guided mode of the waveguide structure (ITO capacitive stack on $Si_3N_4$) is simulated as a function of the applied voltage, which is shown in Fig. 1f.

**ITO deposition and measurement of carrier concentration**

The properties of the deposited ITO films are process dependent and can be optimized for the given application. The ITO films in this work were deposited using DC sputtering[42]. It has been reported that lower sputter deposition power results in a lower as-deposited film carrier concentration[42]. In this work, ITO films at different sputtering powers were deposited. We found that sputtering power below 30 W resulted in very slow deposition speed and poor film quality, therefore the lowest sputtering power level was set to 30 W. The refractive index data for the ITO films were measured using Woollam V-VASE Spectroscopic Ellipsometer. The measured ellipsometry data was first converted to refractive index by fitting to a smoothing spline. This refractive index data was then fit to the Drude-Lorentz model of permittivity[42], which then was used to estimate the carrier concentration of the ITO films, $N_c$ (Fig. 1e). Note that the lower limit of the wavelength range in this measurement was chosen based on the performance of the ellipsometer. An ITO film deposition power of 30 W was used for the fabrication of our devices.

**Measurement of low-speed electro-optic response**

A QPhotonics 637.9 nm diode laser is fiber coupled to the MZM via a grating coupler. The input is split into two paths using 50/50 multimode interference couplers (MMI). The phase modulated optical waves from the two paths are then combined using a 1x2 MMI and coupled to the output fiber via the output grating coupler. The input voltage is applied to the MZM using a GSG RF probe. The output optical signal is photodetected and monitored on an oscilloscope. The $V_\pi$ is extracted by fitting the measured data to the theoretical transfer function of a push-pull MZM (*i.e.* a $\cos^2(x)$ function).

**Measurements of high-speed electro-optic response:** The same components used in the low-speed electro-optic response measurement are used to couple to and from the MZM and apply the modulating voltage. The modulated optical signals were detected using a New Focus 1807-FS silicon balanced optical receiver and monitored on a spectrum analyzer.

**Code availability**

All codes produced during this research are available from the corresponding author upon reasonable request.

**Author contributions**

F. A. and N. M. conceived the idea. N. M. conducted numerical simulations, designed and fabricated the devices and performed the experiments and collected data. F. A. and N. M. reviewed, studied and discussed the experimental and numerical simulation results, and discussed the main outcomes of the project. F. A. directed and supervised the project. N. M. and F.A. wrote the manuscript.

**Competing interests**

The authors declare no competing interests.